\documentclass[conference]{IEEEtran}
\IEEEoverridecommandlockouts

\usepackage{cite}
\usepackage{mwe}
\usepackage{subfig}
\usepackage{setspace}
\usepackage{algorithm}
\usepackage{algorithmic}
\usepackage{adjustbox}
\usepackage{physics}

\usepackage{float}%
\usepackage{commath}
\usepackage{ifpdf}
\usepackage{cuted}
\usepackage{lipsum}
\usepackage{diagbox}
\usepackage{multirow}
\usepackage{romannum}
\usepackage{etoolbox}
\usepackage{mathtools}
\usepackage{amsfonts, amsmath, amsthm, amssymb}
\usepackage{bbm}
\usepackage{epsfig, graphicx, wrapfig, color}
\usepackage[normalem]{ulem}
\usepackage{xcolor}
\usepackage{mathtools}
\mathtoolsset{showonlyrefs}

\usepackage{mathtools}
\mathtoolsset{showonlyrefs}

\usepackage{annotate-equations}
\usepackage[skins,listings]{tcolorbox}

\usetikzlibrary{arrows.meta,positioning}

\usepackage{tikz}
\usetikzlibrary{tikzmark} 
\usepackage{hf-tikz} 

\usepackage{comment}

\DeclareMathOperator{\sgn}{sgn}

\newcommand{\F}{\mathbb{F}}
\newcommand{\R}{\mathbb{R}}

\DeclareMathAlphabet{\mathcal}{OMS}{cmsy}{m}{n}
\SetMathAlphabet{\mathcal}{bold}{OMS}{cmsy}{b}{n}

\allowdisplaybreaks

\usepackage{xcolor}

\definecolor{2stageup}{RGB}{97,108,140} 
\definecolor{2stagedown}{RGB}{178,213, 155}

\usepackage{url}

\def\BibTeX{{\rm B\kern-.05em{\sc i\kern-.025em b}\kern-.08em
    T\kern-.1667em\lower.7ex\hbox{E}\kern-.125emX}}
\begin{document}

\title{Learned codes for broadcast channels with feedback\thanks{The work of N. Devroye and Y. Zhou was partially supported by NSF under awards 2217023,  2240532, and 1900911. The contents of this article are solely the responsibility of the authors and do not necessarily represent the official views of the NSF.}}

\author{\IEEEauthorblockN{Yingyao Zhou$^1$,  Natasha Devroye$^1$}
\IEEEauthorblockA{$^1$University of Illinois Chicago, Chicago, IL, USA}

\IEEEauthorblockA{\{yzhou238, devroye\}@uic.edu}
}


\maketitle

\begin{abstract}
We focus on designing error-correcting codes for the symmetric Gaussian broadcast channel with feedback. Feedback not only expands the capacity region of the broadcast channel but also enhances transmission reliability. In this work, we study the construction of learned finite blocklength codes for broadcast channels with feedback. Learned error-correcting codes, in which both the encoder and decoder are jointly trained,  have shown impressive performance in point-to-point channels, particularly with noisy feedback. However, few learned schemes exist for multi-user channels. Here, we develop a lightweight code for the broadcast channel with feedback that performs well and operates effectively at short blocklengths. 
\end{abstract}


\section{Introduction}
We study the transmission of independent messages over a two-user static Gaussian broadcast channel with either noiseless or noisy feedback, where both the forward and feedback links are corrupted by independent Gaussian noises. While feedback does not increase the capacity of a memoryless point-to-point (P2P) channel \cite{shannon1956zero}, it can expand the capacity region of a broadcast channel even when the receiver noises are independent \cite{ozarow1984achievable}, or with only one user's feedback \cite{bhaskaran2008gaussian}.

For the point-to-point additive white Gaussian noise (AWGN) channel with feedback, several well-known schemes exist: Schalkwijk-Kailath  \cite{schalkwijk1966coding} introduced the ``SK'' scheme, a capacity-achieving linear coding scheme for noiseless feedback that achieves doubly exponential probability of error decay. 
For noisy feedback, Chance and Love proposed the CL scheme, a concatenated coding approach that uses passive feedback \cite{chance2011concatenated}, while  the Modulo-SK scheme introduced in \cite{ben2017interactive} uses active feedback, achieving superior performance.

For the two-user Gaussian broadcast channel with feedback (GBCF),  Ozarow extended the SK scheme to the ``OL'' scheme, which also achieves doubly exponential error decay~\cite{ozarow1984achievable}. 
Additionally, the work in \cite{murin2014ozarow} enhanced the OL scheme by adopting estimators with memory, i.e. that use the last two channel outputs $y_{i-1}, y_i$ rather than simply the current channel output $y_i$, resulting in the ``EOL'' scheme. This enhancement not only improves the achievable rates of the OL scheme but also increases transmission reliability. In addition to these SK-type coding schemes, control-theory-based codes have been developed in for example \cite{ardestanizadeh2012lqg}, where  LQG codes for the K-user GBCF, generalizing the schemes in \cite{elia2004bode} were proposed. 
For the GBCF with noisy feedback, a linear concatenated coding scheme for the K-user GBCF is proposed in \cite{ahmad2015concatenated}.

The focus of this paper is not on achievable rates or error exponents, but rather on constructing practical codes at  \textit{finite} blocklength, evaluating them based on their \textit{probability of error} for a fixed rate.  To achieve low bit or block error rates for finite block lengths, deep learning has successfully been used to design encoder/decoder for P2P error-correcting codes with feedback (DL-ECFCs). These DL-ECFCs~\cite{kim2018deepcode, safavi2021deep, mashhadi2021drf, shao2023attentioncode, ozfatura2022all, kim2023robust, ankireddy2024lightcode} outperform traditional analytically constructed codes in most cases, especially in noisy feedback. DL-ECFCs can be categorized into two types: \textit{bit-by-bit}, where one bit of information is transmitted at a time, and \textit{symbol-by-symbol}, where message bits of length $K$ are mapped to a symbol for transmission, similar to SK-type codes. Bit-by-bit schemes~\cite{kim2018deepcode, safavi2021deep, mashhadi2021drf, shao2023attentioncode} use a two-phase encoding process with a fixed rate of $\frac{1}{3}$, as first introduced in Deepcode~\cite{kim2018deepcode}. Symbol-by-symbol schemes~\cite{ozfatura2022all, kim2023robust, ankireddy2024lightcode} achieve variable rates by adjusting the number of channel uses. The lightweight symbol-by-symbol code in~\cite{ankireddy2024lightcode} achieves state-of-the-art performance for AWGN channels with feedback. However, few deep learning strategies have been applied to multi-user channels; \cite{li2022deep} extends Deepcode to fading Gaussian broadcast channels with feedback, while \cite{ozfatura2023not} focuses on the multiple-access channel (MAC) with feedback. 

{\bf Contribution}: {We generalize Lightcode~\cite{ankireddy2024lightcode} to the symmetric GBCF with independent feedback links\footnote{Our code is available at \url{https://github.com/zyy-cc/GBCF.git}}, achieving superior performance in the short blocklength regime, even with noisy feedback. {Parallel work by~\cite{malayter2024deep} also extends Lightcode to GBCF but follows a different structure. Our scheme uses time-division transmission in the first two rounds, while theirs transmits both messages in the first round.} Additionally, we provide an initial interpretation of our learned codes, revealing an approximately linear relationship between the encoder output and the received feedback.}

{\bf Notation}: Random variables are denoted by capital letters and specific instances by lowercase letters. Vectors are represented in bold, with superscripts indicating their length.  Subscripts $u$ distinguish between users, and $i$ denotes time indices. $\text{SNR}_{u, f}$ and $\text{SNR}_{u, fb}$ represent the Signal-to-Noise Ratios for the forward and feedback channels for user $u$, respectively. Probability is denoted by $\mathbb{P}(\cdot)$ and expectation by $\mathbb{E}(\cdot)$. Real values are represented by $\mathbb{R}$, and $\mathbb{F}_2$ denotes the finite field with elements 0 and 1. Additionally, $\sgn(x) = 1$ if $x \geq 0$, and $0$ otherwise. Moreover, $x \text{ dB} = 10\log{10}(x)$.


\section{System Model}
\begin{figure}[ht]
\vspace{-4mm}
    \centering
\includegraphics[width=0.9\columnwidth]{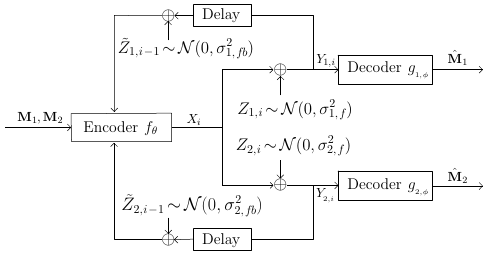}
    \caption{Gaussian broadcast channel with noisy feedback}
    \label{fig:broadcastchannel}
    \vspace{-3mm}
\end{figure}

We consider the real-valued Gaussian Broadcast Channel with feedback for two users (GBCF), as illustrated in Fig. \ref{fig:broadcastchannel}. The transmitter sends independent messages $\mathbf{M}_u = (M_{u,1}, \dots, M_{u,K_u})\in \mathbb{F}_2^{K_u}$ to each user $u \in \{1, 2\}$. A total of $K_1 + K_2$ message bits are sent over $N$ channel uses, at the code rate for each user  $R_u = \frac{K_u}{N}$. At time $i \in \{1, \ldots, N\}$, the received symbols for user $u \in \{1, 2\}$ are given by:
\begin{align}
    Y_{u, i} = X_i + Z_{u,i}
\end{align}
where $X_{i} \in \mathbb{R}$ are the transmitted symbols of the codewords ${\bf X}$ subject to the average power constraint $\frac{1}{N}\mathbb{E}\left[\lVert \mathbf{\mathbf{X}} \rVert_2^2\right]\leq P$, and $Z_{u,i}\sim \mathcal{N}(0,\sigma_{u,f}^2)$ are independent and identically distributed (i.i.d.) Gaussian noises across different users and time.  Each receiver sends the received symbols back to the transmitter through the feedback links with one unit delay. For noiseless feedback, the received feedback equals the channel outputs, $\tilde{Y}_{u, i-1} = Y_{u, i-1}$. However, for noisy feedback, the received feedback is expressed as $\tilde{Y}_{u, i-1} = Y_{u, i-1} + \tilde{Z}_{u,i-1}$ where $\tilde{Z}_{u,i-1}\sim \mathcal{N}(0,\sigma_{u,fb}^2)$ are also i.i.d. Gaussian noises.

In this work, we focus on the symbol-by-symbol coding, similar to the OL scheme, which maps message bits of length $K$ to a symbol and iteratively refines the transmitted symbols. At the $i$-th transmission, the encoding function $f_\theta$ generates the transmitted symbol $X_i$ using the message bits $\mathbf{M}_1$ and $\mathbf{M}_2$, along with the past feedback from both users $\mathbf{\tilde{Y}}_u^{i-1} = \left[\tilde{Y}_{u, 1},\dots,\tilde{Y}_{u, i-1}\right]$, defined as
\begin{equation}
    X_i = c_i f_\theta(\mathbf{M}_1, \mathbf{M}_2, \mathbf{\tilde{Y}}_{1}^{i-1}, \mathbf{\tilde{Y}}_2^{i-1}) 
\end{equation}
where $c_i$ are scaling factors that ensure the power constraint is satisfied.
After $N$ transmissions, the decoder for each user estimates the message bits from their own received noisy codewords, expressed as $\mathbf{\widehat{M}}_u =g_{u,\phi}(\mathbf{Y}_u^{N}) \in \F_2^{K_u}$, where $\mathbf{Y}_u^{N} = \left[Y_{u, 1},\dots,Y_{u, N}\right]$.
The performance metric  is the average block error rate (BLER), defined as $\text{BLER}_u=\mathbb{P}(\mathbf{M}_{u}\neq \widehat{\mathbf{M}}_{u})$ for user $u$. The joint BLER is taken as the average BLER between the two users, given by $\text{BLER}=\frac{1}{2}{(\text{BLER}_1 + \text{BLER}_2)}$. For simplicity, we set $K_1 = K_2$ and focus on the symmetric case where $\sigma_{1, f}^2 = \sigma_{2, f}^2$ and $\sigma_{1, fb}^2 = \sigma_{2, fb}^2$.

Most previous work has focused on analytical linear code design for  the encoder $f_{\theta}$ and the decoders $g_{1, \phi}$ and $g_{2, \phi}$. In this work, we aim to replace them with neural networks parameterized by $\theta$ and $\phi$ to minimize the BLER.


\section{Past Work}
In this section, we provide a brief overview of deep-learned error-correcting feedback codes (DL-ECFCs) for point-to-point AWGN channels with feedback (P2P-AWGN-F) and the analytical linear coding schemes for the GBCF.

\subsection{DL-ECFCs}
{\bf Deepcode}: Deepcode \cite{kim2018deepcode} is the first DL-ECFC for the P2P AWGN channel with passive feedback, achieving a code rate of $\frac{1}{3}$. {The encoder and decoder are based on recurrent neural networks (RNNs) and are trained jointly.} Our previous work \cite{US-ISIT-2024-IEEE,zhou2024higher} explains how the Deepcode encoder utilizes past feedback for error correction and develops an analytical encoder and decoder with significantly fewer learned parameters that achieves comparable performance. The study in \cite{li2022deep} extends Deepcode to the symmetric two-user fading GBCF;  here we consider static channels. GBCF-Deepcode \cite{li2022deep} encoding has two phases: in the first, the BPSK-modulated message bits for both users are transmitted uncoded; in the second, the encoder generates a \textbf{single} parity bit based on feedback from the two receivers. 

{Deepcode achieves a lower bit error rate (BER) than analytical codes with passive feedback but is limited to a code rate of $\frac{1}{3}$. Additionally, for message bits of length $K$, the total number of transmissions—including both forward and backward passes—grows linearly with $K$, potentially causing significant delays.}



{\bf Lightcode}: To address these limitations, \cite{ozfatura2022all} proposes a symbol-by-symbol coding method for P2P-AWGN-F using a transformer-based structure called GBAF, which achieves remarkably low BLER. Lightcode~\cite{ankireddy2024lightcode} eliminates the self-attention and positional encoding components of GBAF, reducing the number of parameters to just $\frac{1}{10}$ of those in GBAF while still outperforming it. We further expand on the structure of Lightcode in Section \ref{sec: lightcode}.




\subsection{Analytical Linear Coding Schemes}
In this subsection, we review the OL and EOL schemes developed for GBCF with \textbf{noiseless} feedback. 

\subsubsection{OL scheme} The OL scheme is an adaptation of the SK scheme \cite{ozarow1984achievable}. Initially, the two message bits, $\mathbf{M}_1$ and $\mathbf{M}_2$, are mapped to real numbers and transmitted separately over the forward channel. The receivers estimate the messages based on the received noisy symbols and send these estimates back to the transmitter via the noiseless feedback link. In subsequent rounds, the transmitter sends a linear combination of the estimation errors made by the two receivers. Each receiver updates its estimate, progressively reducing the errors.


{\bf Initialization}: The message bits $\mathbf{M}_u$, where $u \in \{1, 2\}$, are mapped to a PAM modulation symbol $\Theta_u$ from the constellation $\{\pm{1\eta}, \pm{3\eta}, \dots, \pm{(2^{K_u}-1)\eta}\}$, with $\eta = \sqrt{\frac{3}{2^{2K_u}-1}}$ to satisfy the unity power constraint \footnote{In \cite{ozarow1984achievable}, message bits are mapped to $\Theta_u$, which ranges from $-\frac{1}{2}$ to $\frac{1}{2}$, then scaled by $\sqrt{12P}$. When $2^{K_u}$ is large, $\mathbb{E}[\Theta^2] \approx \frac{1}{12}$. Since our $K_u$ is small, we use the method described in our work.}. The estimated message at receiver $u$ at time $i$ is denoted by $\widehat{\Theta}_{u,i}$, with the estimation error $\epsilon_{u,i} = \widehat{\Theta}_{u,i} - \Theta_u$. The mean square error is represented by $\alpha_{u,i} = \mathbb{E}[\epsilon_{u,i}^2]$, and the correlation coefficient between the estimation errors of the two receivers is defined as $\rho_i = \frac{\mathbb{E}[\epsilon_{1,i} \epsilon_{2,i}]}{\sqrt{\alpha_{1,i} \alpha_{2,i}}}$.
In the first two rounds, the transmitter sends with power $P$ as 
\begin{equation}
    X_i = \sqrt{P}\Theta_i, \quad i = 1, 2
\end{equation}
After the first transmission, receiver 1 estimates $\widehat{\Theta}_{1,1} = \widehat{\Theta}_{1,2} = \frac{Y_{1,1}}{\sqrt{P}}$ and ignores the second transmission. Similarly, receiver 2 ignores the first transmission and estimates $\widehat{\Theta}_{2,1} = \widehat{\Theta}_{2,2} = \frac{Y_{2,2}}{\sqrt{P}}$. In this case, $\alpha_{u,2} = \frac{\sigma_{u,f}^2}{P}$ and $\rho_2 = 0$. In this work, we focus on finite channel uses and apply LMMSE estimation in the first two rounds to improve BLER performance \cite{ben2017interactive}. The estimate is given by
\begin{equation}
    \widehat{\Theta}_{u,i} = \frac{\sqrt{P}}{P + \sigma_{u,f}^2} Y_{u,i}, \quad i = 1, 2, \quad u = 1, 2
\end{equation}


{\bf Encoding}: 
For the $i$-th transmission, where $i \geq 3$, the transmitter sends:
\begin{equation}
    X_i = \sqrt{\frac{P}{D_{i-1}}}\left[\frac{\epsilon_{1, i-1}}{\sqrt{\alpha_{1, i-1}}} + \frac{\epsilon_{2, i-1}}{\sqrt{\alpha_{2, i-1}}}g\sgn(\rho_{i-1})\right]
\end{equation}
where $D_{i-1} = 1 + g^2 + 2g\abs{\rho_{i-1}}$, and $g$ is a nonnegative constant that balances the trade-off between the two users. To achieve similar BLER for both users, we assume $g = 1$.

{\bf Decoding}: At the receiver $u$, a memoryless MMSE estimator is used to estimate $\epsilon_{u,i-1}$ based \textbf{only} on $Y_{u,i}$, denoted by $\hat{\epsilon}_{u,i-1} = \mathbb{E}[\epsilon_{u,i-1} | Y_{u,i}]$. The receiver $u$ then updates its estimate as follows:
\begin{equation}
\widehat{\Theta}_{u,i} =   \widehat{\Theta}_{u,i-1} - \frac{\mathbb{E}[\epsilon_{u,i-1}Y_{u,i}]}{\mathbb{E}[Y_{u,i}^2]}Y_{u,i}, \quad u = 1, 2
\end{equation}
where $\mathbb{E}[Y_{u,i}^2] = P + \sigma_{u,f}^2$, $\mathbb{E}[\epsilon_{1,i-1} Y_{1,i}] = \sqrt{\frac{P}{D_{i-1}}} \sqrt{\alpha_{1,i-1}}(1 + g\abs{\rho_{i-1}})$, and $\mathbb{E}[\epsilon_{2,i-1} Y_{2,i}] = \sqrt{\frac{P}{D_{i-1}}} \sqrt{\alpha_{2,i-1}}(g + \abs{\rho_{i-1}})\sgn(\rho_{i-1})$. 

\begin{figure}[htbp]
    \centering
    \includegraphics[width=0.9\columnwidth]{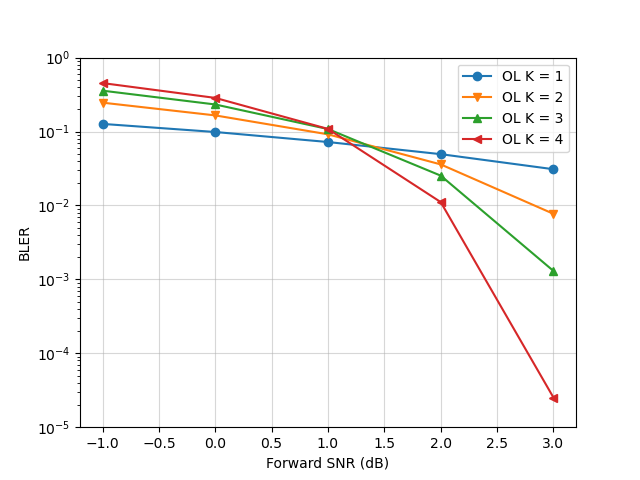}
    \caption{BLER performance for code rate $\frac{1}{3}$ with varying message bit lengths $K$.}
    \label{fig:olsamerate}
    \vspace{-3mm}
\end{figure}

We experimentally observed that for the same code rate of $\frac{1}{3}$, different message bit lengths $K = K_1=K_2$ result in varying BLER performance, as shown in Fig. \ref{fig:olsamerate}. At low $\text{SNR}_f$, where noises significantly affect the signal, shorter message bit lengths perform better due to the increased distance between constellation points. {Conversely, at high $\text{SNR}_f$, longer message bit lengths $K$—and thus more channel uses $N = 3K$—lead to better performance, as the decoding error of the OL schemes decays doubly exponentially in $N$.}

However, $K$ cannot be too large due to precision issues and quantization errors associated with $2^K$ PAM modulation. {The advanced analog-to-digital converters, such as the Texas Instruments ADS1263 and Analog Devices AD4134, support up to 32-bit quantization \cite{shao2023attentioncode}.} In our experiments, when $K \geq 25$, the BLER starts to increase rather than decrease.  To accommodate longer lengths $L$, we recommend selecting an appropriate value for $K$ to avoid precision issues—{using a small $K$ at low SNR and a large $K$ at high SNR}—and then dividing it into $l = \frac{L}{K}$ chunks of bits. Each chunk, with a message bit length of $K$, is processed using the desired coding scheme. The deep-learned codes also adhere to this rule; however, with more rounds, the input space increases, which raises training difficulty.

\subsubsection{EOL scheme}
The EOL scheme \cite{murin2014ozarow} extends the previous OL scheme by incorporating an MMSE estimator based on the current and previous channel outputs rather than just the current output. The transmission process remains nearly the same as before, except that the receiver now estimates $\hat{\epsilon}_{u,i-1} = \mathbb{E}[\epsilon_{u,i-1} | Y_{u,i-1}, Y_{u,i}]$. For further details, we refer readers to \cite{murin2014ozarow}.

\section{GBCF-Lightcode-sep} \label{sec: lightcode}
 {We now extend Lightcode to GBCF-Lightcode-sep, which supports a wider range of code rates. While the concurrent~\cite{malayter2024deep} sends a symbol containing both messages in the first channel use and uses the remaining channel uses ($N-1$) to help the decoder refine this, our scheme uses separate channel uses for each of the messages first, before simultaneously refining in the remaining channel uses ($N-2$). For a fixed blocklength, our scheme thus has less refinement rounds compared to \cite{malayter2024deep} but has perhaps ``clearer'' initial message transmission.}
 
 {Learned codes for GBCF offer superior BLER performance compared to analytical schemes, particularly when the forward SNR is high. They also remain effective with noisy feedback, even when the noise variance in the feedback link is high.
 }



\subsection{Structure}
The structure is inspired by the block coding approach in \cite{ozfatura2022all, ankireddy2024lightcode}, with the design consisting of a feature extractor (FE) and a multilayer perceptron (MLP) module.

\begin{figure*}[ht]
    \centering
    \includegraphics[width=0.8\textwidth]{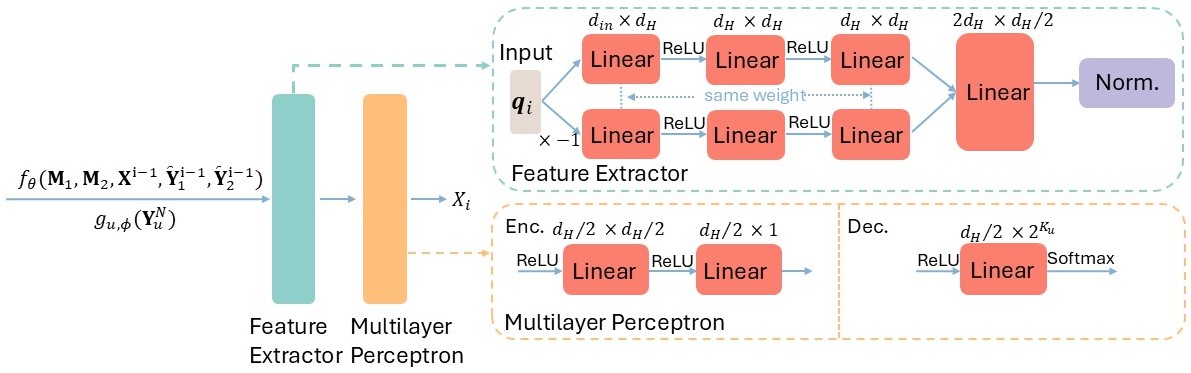}
    \caption{Design of the GBCF-Lightcode-sep (left) and its detailed structure (right).}
    \label{fig:bclight}
    \vspace{-3mm}
\end{figure*}

{\bf Feature extractor}: {The structure is shown in Fig. \ref{fig:bclight}. In \cite{ankireddy2024lightcode}, the original dimension $d_H$ is set to 32. Here, we double the input information and experimentally find that the increased dimension improves performance. Therefore, we set $d_H = 64$.} Layer normalization is added to stabilize training.

{\bf Multilayer perceptron}: The output of the feature extractor is passed to the MLP layers. The encoder uses a two-layer MLP to generate a symbol for each channel use, while the decoder employs a one-layer MLP. For the decoder on the receiver side, $u \in \{1, 2\}$, the index for each message bit $\mathbf{M}_u$ ranges from $0$ to $2^{K_u} - 1$. Thus, the output of the linear layer is projected to a vector of dimension  $C_u = 2^{K_u}$, with a softmax function applied to denote the probability for each index. For a vector $\mathbf{v} \in \R^{C_u}$, the $i$-th element after applying softmax is given by $\text{softmax}(v_{i}) = \frac{e^{v_{i}}}{\sum_{j=1}^{C_u} e^{v_{j}}}.$

\subsection{Transmission}
The entire transmission design is inspired by the OL scheme \cite{ozarow1984achievable}. The learned parameters are highlighted in \textcolor{red}{red}.

{\bf Initialization}: Similar to the OL scheme, in the first two rounds, the message $\mathbf{M}_u \in \mathbb{F}_2^{K_u}$ is mapped to the PAM modulation symbol $\Theta_u$ for $u \in \{1, 2\}$  and transmitted separately.
\begin{equation}
    X_i = \textcolor{red}{\beta_i}\Theta_i, \quad i = 1, 2
\end{equation}


{\bf Encoding}: In the subsequent rounds, the learned encoder generates the transmitted symbols based on the previously generated symbols and the feedback from the two receivers from the last $i-1$ rounds. The generated symbol is given by 
\begin{equation}
    X_i = \textcolor{red}{\beta_i}f_{\textcolor{red}{\theta}}(\mathbf{X}^{i-1}, \mathbf{\tilde{Y}}_1^{i-1}, \mathbf{\tilde{Y}}_2^{i-1}), \quad i = {3, \dots, N}
\end{equation}
where $\mathbf{\tilde{Y}}_{u}^{i-1} = [\tilde{Y}_{u,1}, \dots,\tilde{Y}_{u,i-1}]$ represents the feedback from receiver $u$, with each element given by $\tilde{Y}_{u,i-1} = X_{i-1} + Z_{u, i-1} + \tilde{Z}_{u,i-1}$. The whole transmission is subject to an average power constraint, meaning $\frac{1}{N}\sum_{i=1}^{N} \textcolor{red}{\beta_i}^2 = P$.  
{The encoding process in~\cite{malayter2024deep} follows a different structure from ours. Instead of transmitting the messages separately, the encoder sends messages of length $2K_u$ for both receivers in the first round $X_1 =\textcolor{red}{\beta_i}f_{\textcolor{red}{\theta}}(\mathbf{M}_1, \mathbf{M}_2)$, with subsequent rounds used for error correction $ X_i = \textcolor{red}{\beta_i}f_{\textcolor{red}{\theta}}(\mathbf{M}_1, \mathbf{M}_2, \mathbf{X}^{i-1}, \mathbf{\tilde{Y}}_1^{i-1}, \mathbf{\tilde{Y}}_2^{i-1})$}

{In the following discussion, we refer to our scheme as ``GBCF-Lightcode-sep'' and the scheme in \cite{malayter2024deep} as ``GBCF-Lightcode''.}

{\bf Decoding}: {For each receiver, the decoder estimates its message based on the received noisy codewords, as does ~\cite{malayter2024deep}:
}
\begin{equation}
    \widehat{\mathbf{M}}_u =g_{u, \textcolor{red}{\phi}}(\mathbf{Y}_{u}^{N}).
\end{equation}

The encoder $f_{\textcolor{red}{\theta}}$ and the two decoders $g_{u, \textcolor{red}{\phi}}$ consist of FE and MLP modules, as shown in Fig. \ref{fig:bclight}. The entire process is outlined in Algorithm \ref{trainlight}. The input dimension to the neural networks is fixed, so the $*$ parts will be padded with zeros to maintain the required dimension. The function $H_{ps}$ represents the power constraint block, while $H_{d2b}$ converts the PAM index to its binary representation. In this work, we use the negative log-likelihood (NLL) loss as the objective function:
\begin{equation}
    \ell_u(\mathbf{p}_u, \mathbf{\hat{p}}_u) = -\frac{1}{B}\sum_{i=1}^{B}\left(\sum_{j=0}^{C_u-1}p_{u, ij} \log(\hat{p}_{u, ij})\right)
\end{equation}
Here, $B$ is the batch size, and $C_u = 2^{K_u}$ represents the number of classes (PAM indices) for receiver $u$, and $p_{u, ij}$ is $1$ if class $j$ is the correct class for the $i$-th sample, and $0$ otherwise. $\hat{p}_{u, ij} \in \mathbb{R}$ denotes the predicted probability for class $j$ of the $i$-th sample. Additionally, during training, to prevent one receiver from significantly outperforming the other, we add a regularization term $(\ell_{1} - \ell_{2})^2$ to ensure that the performance of both receivers remains comparable.

\begin{algorithm}
    \caption{GBCF-Lightcode-sep (Encoder $f_{\textcolor{red}{\theta}}$, Decoder 1 $g_{1, \textcolor{red}{\phi}}$, Decoder 2 $g_{2, \textcolor{red}{\phi}}$, Channel Uses $N$, Message Bits $\mathbf{M}_1$, $\mathbf{M}_2$)}
    \label{trainlight}
    \begin{algorithmic}[1]
        \STATE \textbf{Initialization:}
        Map $\mathbf{M}_1$, $\mathbf{M}_2$ to PAM symbols $\Theta_1$, $\Theta_2$; set $X_1 = \beta_1\Theta_1$, $X_2 = \beta_2\Theta_2$
        \FOR {each $i \in [3, N]$}
            \STATE Update encoder input vector: 
            \STATE $\mathbf{q}_{i} = [X_1, X_2, \dots, X_{i-1}, *, \tilde{Y}_{1,1}, \dots, \tilde{Y}_{1,i-1}, *,$
            \STATE \quad \quad \quad $ \tilde{Y}_{2,1}, \dots, \tilde{Y}_{2,i-1}, *]$
            \STATE Generate transmitted symbol: 
            $X_i = \textcolor{red}{\beta_i} H_{ps}[f_{\textcolor{red}{\theta}}(\mathbf{q}_{i})]$
            \STATE Transmit $X_i$ to both receivers
        \ENDFOR
        \STATE Receiver 1: $\mathbf{\hat{p}}_{1} = g_{1, \phi}(Y_{1}^N) \in \mathbb{R}^{C_1}$
        \STATE Receiver 2: $\mathbf{\hat{p}}_{2} = g_{2, \phi}(Y_{2}^N) \in \mathbb{R}^{C_2}$
        \STATE Compute NLL loss: $\ell_{1} + \ell_{2} + (\ell_{1} - \ell_{2})^2$
        \STATE Classification: $I_u = \text{argmax}(\mathbf{\hat{p}}_{u})$, where $u \in \{1, 2\}$
        \STATE Get bit representation: $\widehat{\mathbf{M}}_u = H_{d2b}(I_u)$, where $u \in \{1, 2\}$
    \end{algorithmic}
\end{algorithm}
{The training parameters are listed in Table \ref{params}, where training is performed at the corresponding forward and feedback SNRs.}

\begin{table}[!t]
\vspace{6mm}
\caption{Training parameters for GBCF-Lightcode-sep}
\label{params}
\centering
\begin{tabular}{|l|c|}
\hline
\bfseries Parameters            & \bfseries Values       \\ \hline
Batch size $B$                      & 100,000                \\ \hline
Optimizer                       & AdamW                  \\ \hline
Weight decay                    & 0.01                   \\ \hline
Epochs                          & 120000                   \\ \hline
Initial learning rate           & 0.002                  \\ \hline
Clipping threshold              & 0.5                    \\ \hline
Power $P$                       & 1                      \\ \hline
\end{tabular}
\end{table}

\section{simulations}
In this section, we evaluate the performance of GBCF-Lightcode-sep and compare it with other existing codes.

\subsection{Noiseless Feedback}
\begin{figure}[ht]
\vspace{-4mm}
    \centering
    \includegraphics[width=0.9\columnwidth]{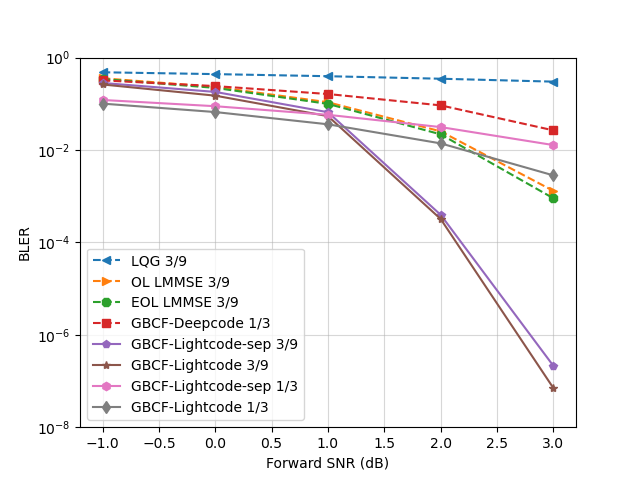}
    \caption{BLER performance across different schemes with noiseless feedback}
    \label{fig:blerperformance}
    \vspace{-1mm}
\end{figure}

{In Fig. \ref{fig:blerperformance} we compare the BLER of codes with a code rate of $\frac{1}{3}$ under noiseless feedback. Although LQG codes outperform OL and EOL asymptotically in terms of achievable rates, LQG demonstrates poorer block error rate (BLER) performance in the finite-length regime. At low SNR\footnote{For a fair comparison, we set the block length to 3 for GBCF-Deepcode.}, shorter messages perform better at the same code rate. As forward SNR increases, the BLER of longer messages with more channel uses decreases significantly. GBCF-Lightcode-sep performs worse than GBCF-Lightcode when $K_u = 1$ but becomes nearly comparable at $K_u = 3$. This suggests that for small message lengths ($K$), transmitting both messages ($\mathbf{M}_1, \mathbf{M}_2$) in the first round for an additional error correction round is more effective, while for larger $K$, time-division in the first two transmissions is preferable.}

\begin{figure}[ht]
    \centering
    \includegraphics[width=0.9\columnwidth]{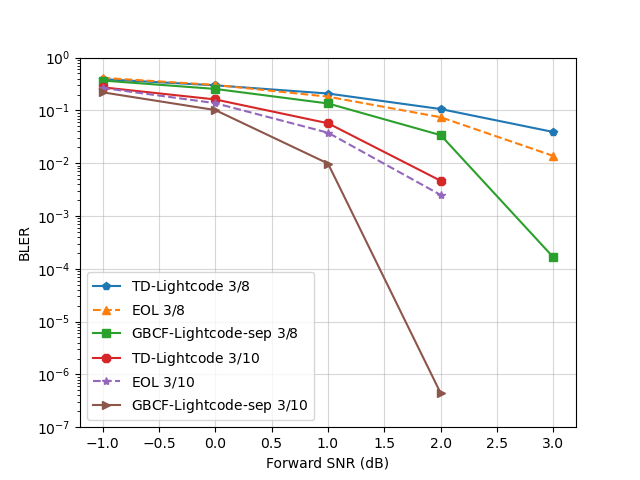}
    \caption{BLER comparison for code rates $\frac{3}{8}$ and $\frac{3}{10}$ with noiseless feedback}
    \label{fig:tdperformance}
    \vspace{-3mm}
\end{figure}

We also consider the Time-Division Lightcode (TD-Lightcode)~\cite{ankireddy2024lightcode} scheme as a benchmark, in which each user transmits their message during an assigned time slot. To ensure fairness between the users, we assume an even number of channel uses, $N$, with $\frac{N}{2}$ time slots allocated for transmitting $\mathbf{M}_1$ and the other $\frac{N}{2}$ slots for transmitting $\mathbf{M}_2$. As shown in Fig. \ref{fig:tdperformance}, the BLER performance is illustrated for code rates $\frac{3}{8}$ and $\frac{3}{10}$ with noiseless feedback. {The results show that, at the same code rate, GBCF codes (including EOL schemes and GBCF-Lightcode-sep) outperform TD-Lightcode, demonstrating that cooperation between the encoder and decoders effectively leverages feedback from both users. Notably, GBCF-Lightcode-sep achieves the best performance among all tested codes. Additionally, the number of channel uses, $N$, is crucial for enhancing BLER performance; even a single additional channel use can lead to a significant improvement.}

\subsection{Noisy Feedback}

\begin{figure}[ht]
    \centering
    \includegraphics[width=0.9\columnwidth]{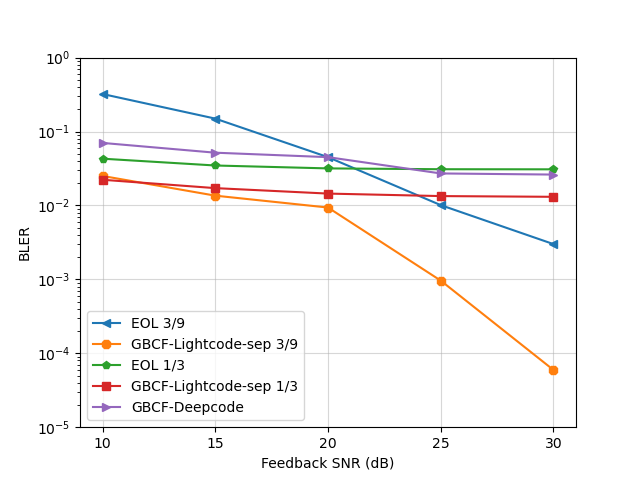}
    \caption{BLER performance with noisy feedback and fixed forward SNR $3$ dB}
    \label{fig:noisyperformance}
    \vspace{-3mm}
\end{figure}

{We also investigate the performance of GBCF-Lightcode-sep under noisy feedback. Fig. \ref{fig:noisyperformance} compares the performance of various schemes, assuming a fixed forward $\text{SNR}_f = 3$ dB. Results indicate that when feedback noise variance is low, symbol-by-symbol coding \footnote{Although the OL and EOL schemes are designed for the noiseless feedback case, here we update the estimation error at the encoder with feedback noise, and apply normalization to ensure the transmitted symbol meets the power constraint.} with a longer block length ($N = 9$) performs well, as additional channel uses allow for better message refinement when feedback is reliable. However, longer block lengths are more sensitive to feedback noise, while shorter block lengths ($N = 3$) are more robust. Notably, GBCF-Lightcode-sep outperforms other codes by adjusting the block length according to feedback noise levels. }



\subsection{Interpretation}
\begin{figure}[ht]
\vspace{-4mm}
    \centering
    \includegraphics[width=\columnwidth]{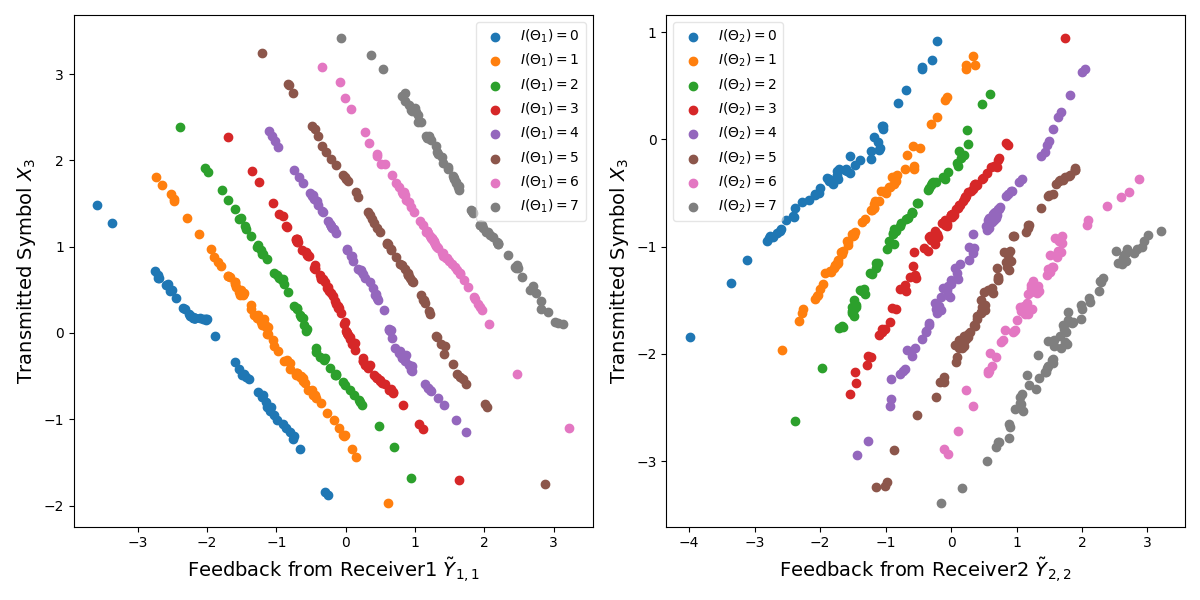}
    \caption{Encoder output at round 3 $X_3$ with respect to the noisy feedback $\tilde{Y}_{1,1}$ (left) and $\tilde{Y}_{2,2}$ (right) at code rate $\frac{3}{9}$, forward SNR $3$ dB and noiseless feedback.}
    \label{fig:interprete}
    \vspace{-3mm}
\end{figure}

In this subsection, we present an initial interpretation of the learned GBCF-Lightcode-sep encoder, examining the relationship between the encoder output at round 3, $X_3$, and past noisy feedback from two receivers, as in \cite{ankireddy2024lightcode}. Here, $I(\Theta_u)$, which ranges from $0$ to $2^{K_u}-1$, represents the PAM index for the message bits of user $u$. For feedback from receiver 1 ($\tilde{Y}_{1,1}$) and receiver 2 ($\tilde{Y}_{2,2}$), we analyze the relationship between $X_3$ and the noisy feedback by setting the message bits and forward noises of the opposite user to $I(\Theta_2) = 0$ and 0 for user 2 (left side of Fig. \ref{fig:interprete}) and $I(\Theta_1) = 0$ and 0 for user 1 (right side of Fig. \ref{fig:interprete}). From Fig. \ref{fig:interprete}, we observe that $X_3$ is approximately a linear function of the feedback channel output, being negatively proportional to $\tilde{Y}_{1,1}$ and positively proportional to $\tilde{Y}_{2,2}$. 



\section{Conclusion}
In this work, we propose a symbol-by-symbol, learned coding scheme for the static Gaussian broadcast channel with feedback. This scheme easily applies to different coding rates and performs especially well at high forward SNR, while remaining robust to noisy feedback. At the same code rate, shorter message lengths (e.g., sending $2^K$ PAM signals with smaller $K$) are preferred at low forward or feedback SNR. Conversely, at high SNR, longer messages (with larger $K$ and correspondingly more channel uses $N$) are more suitable. We also provide an initial interpretation of our scheme, showing that the learned encoder output at round 3 is a linear function of the channel outputs. {Compared to GBCF-Lightcode in \cite{malayter2024deep}, our GBCF-Lightcode-sep uses time-division in the first two rounds rather than simultaneosuly sending one symbol with two messages in the first round. We thus tradeoff one less refinement round (for 2 users) with a clearer initial guess.} Depending on the scenario, the learned GBCF codes achieve the best-known finite blocklength performance for $N \leq 10$.


\bibliographystyle{IEEEtran}
\bibliography{reference}

\begin{thebibliography}{10}
\providecommand{\url}[1]{#1}
\csname url@samestyle\endcsname
\providecommand{\newblock}{\relax}
\providecommand{\bibinfo}[2]{#2}
\providecommand{\BIBentrySTDinterwordspacing}{\spaceskip=0pt\relax}
\providecommand{\BIBentryALTinterwordstretchfactor}{4}
\providecommand{\BIBentryALTinterwordspacing}{\spaceskip=\fontdimen2\font plus
\BIBentryALTinterwordstretchfactor\fontdimen3\font minus \fontdimen4\font\relax}
\providecommand{\BIBforeignlanguage}[2]{{%
\expandafter\ifx\csname l@#1\endcsname\relax
\typeout{** WARNING: IEEEtran.bst: No hyphenation pattern has been}%
\typeout{** loaded for the language `#1'. Using the pattern for}%
\typeout{** the default language instead.}%
\else
\language=\csname l@#1\endcsname
\fi
#2}}
\providecommand{\BIBdecl}{\relax}
\BIBdecl

\bibitem{shannon1956zero}
C.~Shannon, ``The zero error capacity of a noisy channel,'' \emph{IRE Transactions on Information Theory}, vol.~2, no.~3, pp. 8--19, 1956.

\bibitem{ozarow1984achievable}
L.~Ozarow and S.~Leung-Yan-Cheong, ``An achievable region and outer bound for the gaussian broadcast channel with feedback (corresp.),'' \emph{IEEE Transactions on Information Theory}, vol.~30, no.~4, pp. 667--671, 1984.

\bibitem{bhaskaran2008gaussian}
S.~R. Bhaskaran, ``Gaussian broadcast channel with feedback,'' \emph{IEEE Transactions on Information Theory}, vol.~54, no.~11, pp. 5252--5257, 2008.

\bibitem{schalkwijk1966coding}
J.~Schalkwijk and T.~Kailath, ``A coding scheme for additive noise channels with feedback--i: No bandwidth constraint,'' \emph{IEEE Transactions on Information Theory}, vol.~12, no.~2, pp. 172--182, 1966.

\bibitem{chance2011concatenated}
Z.~Chance and D.~J. Love, ``Concatenated coding for the awgn channel with noisy feedback,'' \emph{IEEE Transactions on Information Theory}, vol.~57, no.~10, pp. 6633--6649, 2011.

\bibitem{ben2017interactive}
A.~Ben-Yishai and O.~Shayevitz, ``Interactive schemes for the awgn channel with noisy feedback,'' \emph{IEEE Transactions on Information Theory}, vol.~63, no.~4, pp. 2409--2427, 2017.

\bibitem{murin2014ozarow}
Y.~Murin, Y.~Kaspi, and R.~Dabora, ``On the ozarow-leung scheme for the gaussian broadcast channel with feedback,'' \emph{IEEE Signal Processing Letters}, vol.~22, no.~7, pp. 948--952, 2014.

\bibitem{ardestanizadeh2012lqg}
E.~Ardestanizadeh, P.~Minero, and M.~Franceschetti, ``Lqg control approach to gaussian broadcast channels with feedback,'' \emph{IEEE transactions on information theory}, vol.~58, no.~8, pp. 5267--5278, 2012.

\bibitem{elia2004bode}
N.~Elia, ``When bode meets shannon: Control-oriented feedback communication schemes,'' \emph{IEEE transactions on Automatic Control}, vol.~49, no.~9, pp. 1477--1488, 2004.

\bibitem{ahmad2015concatenated}
Z.~Ahmad, Z.~Chance, D.~J. Love, and C.-C. Wang, ``Concatenated coding using linear schemes for gaussian broadcast channels with noisy channel output feedback,'' \emph{IEEE Transactions on Communications}, vol.~63, no.~11, pp. 4576--4590, 2015.

\bibitem{kim2018deepcode}
H.~Kim, Y.~Jiang, S.~Kannan, S.~Oh, and P.~Viswanath, ``Deepcode: Feedback codes via deep learning,'' \emph{NeurIPS}, vol.~31, 2018.

\bibitem{safavi2021deep}
A.~R. Safavi, A.~G. Perotti, B.~M. Popovic, M.~B. Mashhadi, and D.~Gunduz, ``Deep extended feedback codes,'' \emph{arXiv preprint arXiv:2105.01365}, 2021.

\bibitem{mashhadi2021drf}
M.~B. Mashhadi, D.~Gunduz, A.~Perotti, and B.~Popovic, ``Drf codes: Deep snr-robust feedback codes,'' \emph{arXiv preprint arXiv:2112.11789}, 2021.

\bibitem{shao2023attentioncode}
Y.~Shao, E.~Ozfatura, A.~Perotti, B.~Popovic, and D.~G{\"u}nd{\"u}z, ``Attentioncode: Ultra-reliable feedback codes for short-packet communications,'' \emph{IEEE Transactions on Communications}, 2023.

\bibitem{ozfatura2022all}
E.~Ozfatura, Y.~Shao, A.~G. Perotti, B.~M. Popovi{\'c}, and D.~G{\"u}nd{\"u}z, ``All you need is feedback: Communication with block attention feedback codes,'' \emph{IEEE Journal on Selected Areas in Information Theory}, vol.~3, no.~3, pp. 587--602, 2022.

\bibitem{kim2023robust}
J.~Kim, T.~Kim, D.~Love, and C.~Brinton, ``Robust non-linear feedback coding via power-constrained deep learning,'' \emph{arXiv preprint arXiv:2304.13178}, 2023.

\bibitem{ankireddy2024lightcode}
S.~K. Ankireddy, K.~Narayanan, and H.~Kim, ``Lightcode: Light analytical and neural codes for channels with feedback,'' \emph{IEEE Journal on Selected Areas in Communications}, pp. 1--1, 2025.

\bibitem{li2022deep}
S.~Li, D.~Tuninetti, and N.~Devroye, ``Deep learning-aided coding for the fading broadcast channel with feedback,'' in \emph{ICC 2022-IEEE International Conference on Communications}.\hskip 1em plus 0.5em minus 0.4em\relax IEEE, 2022, pp. 3874--3879.

\bibitem{ozfatura2023not}
E.~Ozfatura, C.~Bian, and D.~G{\"u}nd{\"u}z, ``Do not interfere but cooperate: A fully learnable code design for multi-access channels with feedback,'' in \emph{2023 12th International Symposium on Topics in Coding (ISTC)}.\hskip 1em plus 0.5em minus 0.4em\relax IEEE, 2023, pp. 1--5.

\bibitem{malayter2024deep}
J.~Malayter, C.~Brinton, and D.~Love, ``Deep learning aided broadcast codes with feedback,'' \emph{arXiv preprint arXiv:2410.17404}, 2024.

\bibitem{US-ISIT-2024-IEEE}
Y.~Zhou, N.~Devroye, G.~Turán, and M.~Žefran, ``Interpreting deepcode, a learned feedback code,'' in \emph{2024 IEEE International Symposium on Information Theory (ISIT)}, 2024, pp. 1403--1408.

\bibitem{zhou2024higher}
Y.~Zhou, N.~Devroye, G.~Turan, and M.~Zefran, ``Higher-order interpretations of deepcode, a learned feedback code,'' in \emph{2024 60th Annual Allerton Conference on Communication, Control, and Computing}, 2024, pp. 1--8.

\end{thebibliography}

\end{document}